\let\includefigures=\iftrue   
\input harvmac 

% Figure definitions 

\input epsf 

\newcount\figno 
\figno=0 
\def\fig#1#2#3{ 
\par\begingroup\parindent=0pt\leftskip=1cm\rightskip=1cm\parindent=0pt 
\baselineskip=11pt 
\global\advance\figno by 1 
\midinsert 
\epsfxsize=#3 
\centerline{\epsfbox{#2}} 
\vskip 12pt 
{\bf Figure \the\figno:} #1\par 
\endinsert\endgroup\par 
} 
\def\figlabel#1{\xdef#1{\the\figno}} 

% Other definitions 

\noblackbox 
\def\IZ{\relax\ifmmode\mathchoice 
{\hbox{\cmss Z\kern-.4em Z}}{\hbox{\cmss Z\kern-.4em Z}} 
{\lower.9pt\hbox{\cmsss Z\kern-.4em Z}} {\lower1.2pt\hbox{\cmsss 
Z\kern-.4em Z}}\else{\cmss Z\kern-.4em Z}\fi}

\font\cmss=cmss10 \font\cmsss=cmss10 at 7pt 
\def\IR{\relax{\rm I\kern-.18em R}} 
 
\def\frac#1#2{{#1 \over #2}}

%%%%%%%%%%%%%%%%%%%%%%%%%%%%%%%%%%%%%%%%%%%%%%%%%%%%%%%%%%%%%%%%% 

%\draftmode   
\def\journal#1&#2(#3){\unskip, \sl #1\ \bf #2 \rm(19#3) }   
\def\andjournal#1&#2(#3){\sl #1~\bf #2 \rm (19#3) }

%   
%%%%%%%%%%%%%%%%%%%%%%%%%%%%%%%%%%%%   
%   

%   
\catcode`\@=11   
\def\slash#1{\mathord{\mathpalette\c@ncel{#1}}}   
\overfullrule=0pt   
   
\def\BB{{\cal B}}

\def\MM{{\cal M}}   
   
\def\OO{{\cal O}}

\def\SS{{\cal S}}

\def\underrel#1\over#2{\mathrel{\mathop{\kern\z@#1}\limits_{#2}}}

\catcode`\@=12   
   
%%%%%%%%%%%%%%%%%%%%%%%%%%%%%%%%%%%%%%%%%%%%%%%%%%%%%%%%%%%%%%   
   
%   

%%%%%%%%%%%%%%%%%%%%%%%%%%%%%%%%%%%%%%%%%%%%%%%%%%%%%%%%%%%%%%   
% new defs:   

\def\myTitle#1#2{\nopagenumbers\abstractfont\hsize=\hstitle\rightline{#1}% 
\vskip 0.5in\centerline{\titlefont #2}\abstractfont\vskip .5in\pageno=0} 

\myTitle{\vbox{\baselineskip12pt\hbox{} 
\hbox{RI-3-02} 
}} {\vbox{ 
        \centerline{A Note on Multi-trace Deformations and AdS/CFT} 
        \medskip 
        }} 
%\medskip 
\centerline{Amit Sever\foot{E-mail : {\tt asever@cc.huji.ac.il.}} and 
Assaf Shomer\foot{E-mail : {\tt shomer@cc.huji.ac.il.}}} 
\medskip 
\centerline{Racah Institute of Physics, The Hebrew University, 
Jerusalem 91904, Israel} 

\bigskip 
%\bigskip 
\noindent 

We derive the general formula, at a finite cutoff, for the change in the 
boundary condition of a scalar field in AdS under a Multiple-trace deformation 
of the dual CFT. Our analysis suggests that fluctuations around the classical 
solution in $AdS$ should not be constrained by boundary conditions.

\Date{March 2002} 

\baselineskip=16pt

%\rightline{CERN/TH-2000-xxx}   
%\rightline{RI-3-02}   
%\Title{   
%\rightline{hep-th/0203168}}   
%{\vbox{\centerline{A Note on Multiple trace Deformations and AdS/CFT.}}}   
%\medskip   
%\centerline{\it Amit Sever and Assaf Shomer}   
%\bigskip   
%\centerline{Racah Institute of Physics, The Hebrew University}   
%\centerline{Jerusalem 91904, Israel}   
%\bigskip\bigskip\bigskip   
%\noindent   
   
\lref\witten{E.~Witten,``Anti-de Sitter space and holography,'' 
Adv.\ Theor.\ Math.\ Phys.\  {\bf 2} (1998) 253 ; hep-th/9802150.} 

\lref\ma{J.~Maldacena, 
``The large $N$ limit of superconformal field theories and supergravity,'' 
Adv.\ Theor.\ Math.\ Phys.\  {\bf 2} (1998) 231 ; 
Int.\ J.\ Theor.\ Phys.\  {\bf 38} (1998) 1113 ; hep-th/9711200.} 

\lref\kw{I.~R.~Klebanov and E.~Witten, 
``AdS/CFT correspondence and symmetry breaking,'' 
Nucl.\ Phys.\ B {\bf 556} (1999) 89 ; hep-th/9905104.} 

\lref\abs{O.~Aharony, M.~Berkooz and E.~Silverstein, 
`Multiple-trace operators and non-local string theories,'' 
JHEP {\bf 0108} (2001) 006 ; hep-th/0105309.} 

\lref\gkp{S.~S.~Gubser, I.~R.~Klebanov and A.~M.~Polyakov, 
``Gauge theory correlators from non-critical string theory,'' 
Phys.\ Lett.\ B {\bf 428} (1998) 105 ; hep-th/9802109.} 

\lref\witmul{E.~Witten, 
``Multi-trace operators, boundary conditions, and AdS/CFT correspondence,'' ;
hep-th/0112258.} 

\lref\agmoo{O.~Aharony, S.~S.~Gubser, J.~Maldacena, H.~Ooguri and Y.~Oz, 
``Large N field theories, string theory and gravity,'' 
Phys.\ Rept.\  {\bf 323} (2000) 183 ; hep-th/9905111.} 

\lref\bss{M.~Berkooz, A.~Sever and A.~Shomer, 
``Double-trace deformations, boundary conditions and spacetime   
singularities,'' ; hep-th/0112264.} 

\lref\freedman{D.~Z.~Freedman, S.~D.~Mathur, A.~Matusis and L.~Rastelli, 
``Correlation functions in the CFT($d$)/AdS($d+1$) correspondence,'' 
Nucl.\ Phys.\ B {\bf 546} (1999) 96 ; hep-th/9804058.} 

\lref\abstwo{O.~Aharony, M.~Berkooz and E.~Silverstein, 
``Non-local string theories on AdS(3) x S**3 and stable  non-supersymmetric 
backgrounds,'' ; hep-th/0112178.} 

\lref\minriv{P.~Minces and V.~O.~Rivelles, 
``Energy and the AdS/CFT correspondence,'' 
JHEP {\bf 0112} (2001) 010 ; hep-th/0110189.} 

\lref\fdrev{E.~D'Hoker and D.~Z.~Freedman,
``Supersymmetric gauge theories and the AdS/CFT correspondence,'' ;
hep-th/0201253.}

\lref\bfb{P.~Breitenlohner and D.~Z.~Freedman,
``Stability In Gauged Extended Supergravity,''
Annals Phys.\  {\bf 144} (1982) 249.}

\lref\muck{W.~Muck,
``An improved correspondence formula for AdS/CFT with multi-trace  operators,''
Phys.\ Lett.\ B {\bf 531} (2002) 301 ; hep-th/0201100.}

\lref\minc{P.~Minces,
``Multi-trace operators and the generalized AdS/CFT prescription,'' ;
hep-th/0201172.}

\newsec{Introduction.} 

The AdS/CFT correspondence \refs{\ma,\witten,\gkp} claims that string theory on 
$AdS_{d+1}  \times \MM_{9-d}$ is dual to a $CFT_d$ that ``lives'' 
on the $d-$dimensional boundary of $AdS_{d+1}.$ 
CFT operators that are dual to single particle states in string 
theory/gravity are usually referred to (in a language borrowed from 
4-dimensional gauge theories) as ``Single-trace'' operators. 
Deformations of the $CFT$ Lagrangian by such Single-trace operators 
are known to be related to deforming the dual string worldsheet by the 
corresponding vertex-operator. In the low energy gravity approximation this 
amounts to changing the vacuum expectation value of the dual gravity field (see 
\refs{\agmoo,\fdrev} for review and references). 
Deforming the $CFT$ Lagrangian by ``Multi-trace'' operators, corresponding to 
multi-particle states in the dual string/gravity picture, was recently 
argued \refs{\abs,\abstwo} to give rise to a non-local generalization of the 
standard string theory worldsheet Lagrangian (NLST). In \bss\ the effect of 
a ``Double-trace'' deformation was analyzed in the gravity approximation 
and was shown to 
give rise to a change in the boundary condition to which the dual gravity 
field was subjected. 
An ansatz for the general case of ``Multi-trace'' deformations was 
given in \witmul\ leading to the same conclusion. 
In this note we derive a general formula, at a finite cutoff, for the change in 
the boundary condition of a scalar field in AdS under a Multiple-trace 
deformation of the 
dual CFT. This is done using a formulation that is not the one 
usually used in the literature. 
Our result agrees with the one suggested in \witmul\ when one removes 
the IR cutoff in AdS. 
The analysis we present raises the question whether or not one should impose 
boundary conditions in $AdS/CFT$. This point is discussed in section 4. Recent 
papers dealing with similar issues are \refs{\minriv,\minc,\muck}.

\newsec{A finite cutoff formulation of the $AdS/CFT$ correspondence.} 

In this section we summarize a finite cutoff formulation of the 
$AdS_{d+1}/CFT_d$ correspondence that was presented in \bss\ and which will be 
used in this note. This formulation 
is convenient for the discussion of Multi-trace deformations. 
We work in Euclidean $AdS_{d+1}$ in the Poincare patch with 
coordinates $(z,\vec{x})$ where the metric is given by (we set $R_{AdS}=1$) 
\eqn\metric{ds^2={dz^2+dx_idx_j\delta^{ij} \over z^2}.} 
The boundary of Euclidean $AdS$ in these coordinates is the surface 
$\partial (AdS) \equiv \{ (z=0,\vec{x}) \} \cup \{ (z=\infty, 
|\vec{x}|=\infty) \}$. 
We introduce a finite IR cutoff in $AdS$ by considering the surface 
$z=\epsilon <<1$. This introduces a problem near the special point in 
this parameterization $(z=\infty,|\vec{x}|=\infty)$ since as we 
approach $|\vec{x}| \rightarrow \infty $ the surface $z=\epsilon$ approaches 
the boundary (see figure 1). 
This is an artifact of our choice 
of coordinates. To avoid this problem we restrict all sources away from 
$\infty$. Taking that into account we 
denote the boundary of $AdS$ space at finite cutoff 
by $\partial \equiv \{ (z=\epsilon,\vec{x}) \}$ and the bulk by 
$\BB \equiv \{ (z>\epsilon,\vec{x}) \}$. 

\fig{AdS space in Poincare coordinates.}{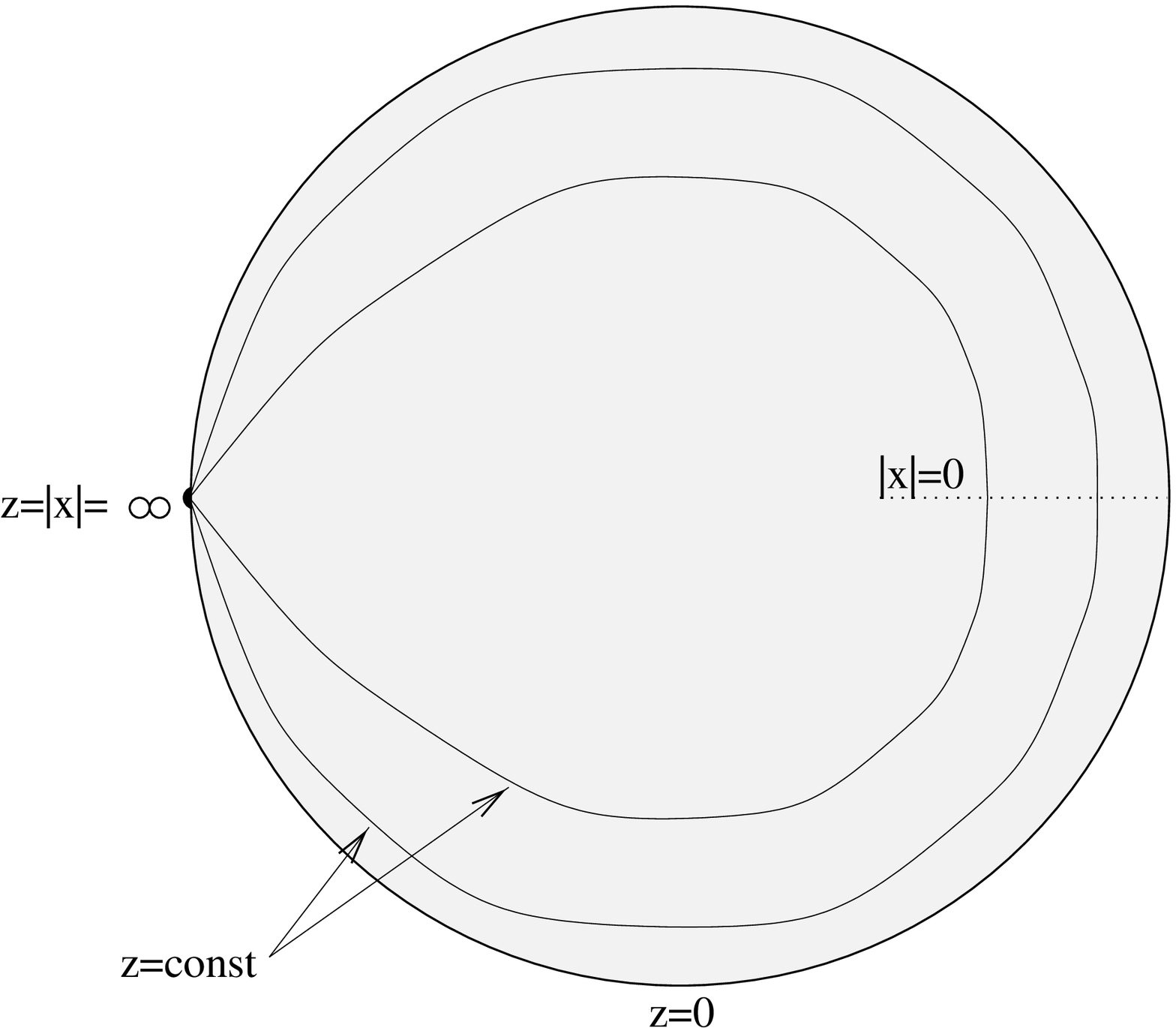}{6 truecm} 
\figlabel\sphrhrzn 

We focus on the case of a scalar field in $AdS$, with mass $m^2 \geq 
-{d^2 \over 4}$, obeying the Breitenlohner Freedman bound and dual to a scalar 
CFT operator of dimension $\Delta$ where $\Delta(\Delta-d)=m^2$. 
We consider the following covariant gravity action which contains a boundary 
term that will prove important for the discussion of boundary conditions: 
\eqn\actmin{\SS_0[\phi]={1 \over 2}\int_{\BB} d^{d+1}x 
\sqrt{g}[g^{\mu\nu}\partial_{\mu}\phi\partial_{\nu}\phi +m^2\phi^2] 
+{\Delta \over 2}\int_{\partial}\sqrt{\widehat{g}}\phi^2,} where 
$\widehat{g}$ stands for the metric restricted to the boundary. In Poincare 
coordinates the expression is: 
\eqn\acgc{\SS_0[\phi]={1 \over 2}\int_{\BB} 
d^dxdzz^{-d+1}[(\partial_z\phi)^2+(\partial_i\phi)^2+{m^2 \over 
z^2}\phi^2]+{\Delta \over 2}\int_{\partial}\epsilon^{-d}\phi^2 .} 
The linear variation of this action is: 
\eqn\linvar{\delta\SS_0[\phi]=\int_{\BB} d^{d+1}x 
\sqrt{g}\delta\phi[-\nabla^2 +m^2]\phi-\int_{\partial} 
\epsilon^{-d}\delta\phi(z\partial_z-\Delta)\phi.} 
In order to have classical solutions there are two choices. Either the field 
obeys a Dirichlet boundary condition $\delta \phi |_{\partial} =0$, or it has to 
satisfy: 
\eqn\mixed{(z\partial_z-\Delta)\phi |_{\partial}=0.} This relation can be 
treated 
either as a boundary condition or as a boundary equation of motion without a 
boundary condition\foot{In the classical approximation the two are the same.}. 
We adopt \mixed\ in what follows and since the above mentioned ambiguity will be 
important later we will refer to \mixed\ as a ``boundary relation". In section 4 
this issue will be discussed. 

The solution to the bulk equation of motion $(\nabla^2 -m^2)\phi=0$ is of the 
form \witten: 
\eqn\clasol{\phi(\vec{x},z)=\int d^dx'\alpha(\vec{x}') { z^{\Delta} \over 
(z^2+|\vec{x}-\vec{x}'|^2)^{\Delta}},} 
where ${ z^{\Delta} \over (z^2+|\vec{x}-\vec{x}'|^2)^{\Delta}}$ is the 
``bulk-boundary'' propagator. 
The meaning of \mixed\ becomes clear when we 
consider the asymptotics of the classical solution near the boundary 
$z \rightarrow 0$. 
\eqn\asympt{\phi(\vec{x},z) =  \left[\alpha(\vec{x})z^{d-\Delta}+\beta(\vec{x}) 
z^{\Delta} \right] (1+\OO(z^2)).} 
In order to get a finite result when $z \rightarrow 0$ we look at: 
\eqn\lookat{z^{\Delta-d}(z\partial_z-\Delta)\phi(\vec{x},z) \rightarrow 
(d-2\Delta)\alpha(\vec{x})+0\times\beta(\vec{x})z^{2\Delta-d}+\dots.} 
The leading term is proportional to the boundary function $\alpha$ and 
demanding $z^{\Delta-d}(z\partial_z-\Delta)\phi(\vec{x},z)=0$ amounts thus to 
the trivial solution $\phi(\vec{x},z)=0.$ 
To have a non trivial solution in this formulation one needs to change the 
action and correspondingly change \mixed. 
The way to do that is to check what boundary relation is obeyed by a general 
solution of the form \clasol: 
\eqn\bcofuy{z^{\Delta-d}(z\partial_z-\Delta)\phi (\vec{x},z)=-2\Delta 
\int d^dx' \alpha(\vec{x}'){z^{2\Delta-d+2} \over (z^2+|\vec{x}-\vec{x}'|^2)^ 
{\Delta+1}}\equiv ({\bf A}\alpha)(\vec{x}),} 
where for brevity of our notation we denote this linear functional acting on 
$\alpha(\vec{x}')$ by $ {\bf A}\left[\alpha(\vec{x}') \right] (\vec{x})$. 
This cumbersome looking integral turns out to have very desirable properties 
as we will demonstrate shortly. 
In order to fix the boundary relation determined by some classical 
function $\rho(\vec{x})$ we thus write: 
\eqn\boucon{z^{\Delta-d}(z\partial_z-\Delta)\phi (\vec{x},z)|_{\partial} 
=({\bf A}\rho)(\vec{x}).} 
This fixes $\alpha(\vec{x})=\rho(\vec{x})$ and allows us to formulate 
conveniently the fact that the boundary relation sets the leading term in 
the expansion of the gravity field near the boundary of AdS 
\refs{\witten,\gkp}\foot{The sub leading term, 
$\beta$ parameterizes the expectation value of the dual CFT field $\OO$ in the 
presence of a source term $\int\rho\OO$. Regularity of the solution in 
Euclidean $AdS$ fixes this term as a function of the leading term $\alpha$.}. 

Moreover, as pointed out in \lookat, it is clear that \boucon\ does not restrict 
terms which scale with $z$ like the subleading term 
$\beta(\vec{x})z^{\Delta}$. 
In the Lorentzian version of the correspondence terms of this type are 
identified with the fluctuating $CFT$ degrees of freedom and it is thus 
desirable 
that they should not be constrained. 

Note that, as usual in field theory, \actmin\ subjected to \boucon\ will not 
have nontrivial classical solutions unless we add to it a boundary source term 
to ensure that its linear variation vanishes also when $\rho \neq 0$. 
The action one gets in the presence of a source is thus: 
\eqn\bndrsrc{\SS_{gr}[\phi ; \rho]\equiv\SS_0[\phi]+ 
\int_{\partial}d^dxd^dx'\epsilon^{-\Delta} 
\phi (\vec{x},z){\bf 
A}(\vec{x},\vec{x}')\rho(\vec{x}')\equiv\SS_0[\phi]+\epsilon^{-\Delta}\phi{\bf 
A}\rho,} 
since the vanishing of the linear variation: 
\eqn\linro{\delta\SS_{gr}[\phi ; \rho]=\int_{\BB} d^{d+1}x 
\sqrt{g}\delta\phi[-\nabla^2 +m^2]\phi-\int_{\partial} 
\epsilon^{-\Delta}\delta\phi [z^{\Delta-d}(z\partial_z-\Delta)\phi- 
{\bf A}\rho],} 
leads either to Dirichlet boundary condition or to the ones we use, 
namely \boucon\foot{Using \boucon\ in \bndrsrc\ one gets the action we used in 
\bss. We will demonstrate below that the general analysis presented here 
reproduces the result in \bss. 
Note that the two actions are not equivalent if we choose 
Dirichlet boundary conditions.}. 

The formulation of the $AdS/CFT$ correspondence \refs{\witten,\gkp} in terms of 
generating functionals in the classical gravity approximation is thus given by: 
\eqn\adscft{Z[\rho]_{CFT} \equiv <e^{\int \rho\OO}>_{CFT}=Z[\rho]_{string} 
\simeq 
\int D[\phi]e^{-\SS_{gr}[\phi; \rho]} \sim e^{-\SS_{gr}[\phi_{\rho}; \rho]},} 
where $<...>_{CFT}$ means the path integral in the CFT, the field $\phi$ is the 
gravity field dual to the CFT 
operator $\OO$, and $\phi_{\rho}$ is the classical solution. 
Since we are interested only in the dynamics of one scalar field we slightly 
abused notations as if it is a field theory path 
integral over this scalar field. This is justified at low enough 
energies before quantum gravity and stringy effects set in, which is 
the regime we are interested in. 
Note that there are two ways to interpret this gravity path integral. Either the 
measure includes \boucon\ as a boundary condition or that there is no boundary 
condition and \boucon\ is imposed at the classical saddle point as a boundary 
equation of motion. 

Let us first show that this reproduces the known results of $AdS/CFT$. 
Evaluating the action \bndrsrc\ on the classical solution \clasol\ obeying 
\boucon, i.e. on 
\eqn\clasro{\phi_{\rho}(\vec{x},z)=\int d^dx'\rho(\vec{x}') { z^{\Delta} \over 
(z^2+|\vec{x}-\vec{x}'|^2)^{\Delta}},} one gets at a finite cutoff $\epsilon$ 
\eqn\clasac{\SS^{\epsilon}[\phi_{\rho}; \rho]={1 \over 2} 
\int_{\partial}d^dx_1d^dx_2d^dx_3{ \rho(\vec{x}_1){\bf A} 
(\vec{x_2},\vec{x}_3)\rho(\vec{x}_3) \over 
(\epsilon^2+|\vec{x}_1-\vec{x}_2|^2)^{\Delta}}.} 
To remove the cutoff we use the formula (for $\gamma > {d \over 2}$):
\eqn\deltf{\lim_{z \rightarrow 0}{z^{2\gamma-d} \over 
(z^2+|\vec{x}-\vec{x}'|^2)^{\gamma}}=\pi^{d \over2}{\Gamma(\gamma-{d 
\over2}) \over \Gamma(\gamma)}\delta^d(\vec{x}-\vec{x}').} 
This gives (see \bcofuy\ for the definition of {\bf A}): 
\eqn\leema{\lim_{z \rightarrow 0}{\bf A}(\vec{x},\vec{x}')= 
\left(-2\pi^{d \over2}(\Delta-{d \over2}){\Gamma(\Delta-{d \over2}) \over 
\Gamma(\Delta)}\right)\delta^d(\vec{x}-\vec{x}').} 
Thus after removing the cutoff the action becomes: 
\eqn\finac{\SS[\phi_{\rho};\rho]\equiv\lim_{\epsilon \rightarrow 
0}\SS^{\epsilon}[\phi_{\rho};\rho]= 
-(\Delta-{d \over 2})\pi^{d \over2} 
{\Gamma(\Delta-{d \over2}) \over \Gamma(\Delta)}\int d^dxd^dx'{\rho(\vec{x}) 
\rho(\vec{x}') \over |\vec{x}-\vec{x}'|^{2\Delta}}.} 
Here we see the desirable properties of this formulation. The specific form of 
${\bf A}$ gave, upon the removal of the cutoff, exactly the correct coefficient 
$-(\Delta-{d \over 2})$ that was found in \freedman\ to be necessary in order 
to obey Ward identities. The minus sign is important for the positivity of the 
$CFT$ two-point function. 

In the marginal case $\Delta={d \over 2}$ the asymptotic behavior is given 
instead of \asympt\ by: 
\eqn\asympm{\phi(\vec{x},z) =  \left[\alpha(\vec{x})z^{d \over 2} 
log({z \over z_0})+\beta(\vec{x})z^{d \over 2} \right] (1+\OO(z^2)),} with $z_0$ 
a constant. 
An elegant property of this formulation is that due to the 
shift of the exponent in ${\bf A}$ with respect to \clasol\ $(\gamma=\Delta+1)$ 
the above procedure remains true also in this marginal case\foot{There is no 
divergence in \leema\ in the marginal case as can be seen by using 
the defining property of the Gamma function $0\times\Gamma(0)=\Gamma(1)=1$.}, 
with the result 
\eqn\leemb{\lim_{z \rightarrow 0}{\bf A}(\vec{x},\vec{x}')= 
{-d\pi^{d \over 2} \over \Gamma({d \over 2}+1)}\delta^d(\vec{x}-\vec{x}').} 
The action one gets in this case is given in \bss. 
Moreover, scalar fields in $d-$dimensions obey a unitarity bound $\Delta \geq 
{d-2 \over 2}$. In this formalism we see this bound to arise naturally as a 
bound on 
the applicability of the derivation leading to \finac. 
Indeed unless $\Delta >{d-2 \over 2}$ the coefficient in front of \leema\ 
diverges. The fact that in ${\bf A}$ the exponent is $\gamma =\Delta +1$ 
captures this 
property\foot{This bound is also necessary for the expansion \asympt\ to give 
the leading terms when $z \rightarrow 0$. Indeed the $\OO(z^2)$ 
corrections are smaller when $\Delta < {d \over 2}$ if 
$d-\Delta<\Delta+2$.}$^,$\foot{The 
case $\Delta={d-2 \over 2}$ corresponds to a free CFT field. It is thus not 
surprising that we do not see this behavior in the gravity approximation.}. 
This formalism is applicable 
also in the range ${d \over 2} >\Delta > {d-2 \over 2}$, however in this range 
there are further subtelties \refs{\kw,\bfb}. 

To summarize, this formalism reproduces correctly the usual $AdS/CFT$ 
results including some of its subtleties. 
It differs from the usual treatment of $AdS/CFT$ in that we use \boucon\ instead 
of Dirichlet boundary condition. In this treatment the CFT source $\rho$ is also 
a source to the gravity field $\phi$ albeit with coupling only on the boundary 
of $AdS$. 
In the next section we show how to generalize to 
the case of multiple trace deformations of the dual $CFT$.

\newsec{AdS/CFT deformed by a Multiple-trace Operator.} 

In this section we derive the effect of deforming the dual CFT with a 
Multi-trace operator. 
Start with an $AdS/CFT$ dual pair and denote the action of the CFT by 
$I_0[\varphi_j]$, where $\varphi_j$ are the fundamental fields of the CFT. 
Now consider, following \witmul, a general deformation of the CFT action 
\eqn\dfmcft{I_0 \rightarrow I_W \equiv I_0+W[\hat{\OO} (\varphi_j)],} by adding 
some function of the Single-trace operator 
\eqn\sigtr{\hat{\OO}\equiv Tr(\prod_j \varphi_j).} 
Expanding the notation of \adscft\ we write the classical gravity approximation 
to the $AdS/CFT$ correspondence as: 
\eqn\adscfb{Z[\rho]_{CFT}\equiv <e^{\int \rho \hat{\OO}}>_{CFT}= 
\int D[\varphi_j] 
e^{-I_0[\varphi_j]}e^{\rho O(\varphi_i)}\simeq \int D[\phi]e^{-\SS_{gr}[\phi; 
\rho]} \sim e^{-\SS_{gr}[\phi_{\rho}; \rho]} ,} 
where as before $\rho$ is a 
classical source, $<\dots>_{CFT}$ stands for $\int 
D[\varphi_j]e^{-I_0[\varphi_j]}(\dots)$ 
and we use the notation $\hat{\OO}$ for the quantum operator 
acting in the CFT Hilbert space and $O$ for the usual (commutative) function 
which enters in the corresponding path integral. 
To find the gravity action dual to the deformed CFT we proceed as follows: 
\eqn\drvtn{\eqalign{Z_{CFT}^W[\rho] 
&\equiv <e^{\int \rho\hat{\OO}}>_{CFT}^W\equiv<e^{\int 
\rho\hat{\OO}-W[\hat{\OO}]}>_{CFT}=\cr & 
=\int D[\varphi_j] e^{-I_0[\varphi_i]-W[O(\varphi_i)]+\rho O(\varphi_i)} 
=\int D[\varphi_i] e^{-I_0}e^{-W[O]}e^{\rho O}=\cr & 
=\int D[\varphi_i] e^{-I_0}e^{-W[{\delta \over \delta \rho}]}e^{\rho O} 
=e^{-W[{\delta \over \delta \rho}]}\int D[\varphi_i] e^{-I_0}e^{\rho O}=\cr & 
=e^{-W[{\delta \over \delta \rho}]}<e^{\int\rho\hat{\OO}}>_{CFT} 
=e^{-W[{\delta \over \delta \rho}]}\int D[\phi]e^{-\SS_{gr}[\phi ; \rho]}=\cr & 
=e^{-W[{\delta \over \delta \rho}]}\int D[\phi]e^{-\left( \SS_0[\phi]+ 
\epsilon^{-\Delta}\phi {\bf A}\rho\right) } 
=\int D[\phi]e^{-\SS_0[\phi]}e^{-W[{\delta \over \delta \rho}]} 
e^{-\epsilon^{-\Delta}\phi{\bf A}\rho}=\cr & 
=\int D[\phi]e^{-\SS_0[\phi]}e^{-W[-\epsilon^{-\Delta}{\bf A} \phi]}e^{- 
\epsilon^{-\Delta}\phi{\bf A}\rho} 
=\int D[\phi]e^{-\SS_{gr}[\phi ; \rho]}e^{-W[-\epsilon^{-\Delta} 
{\bf A}\phi]} \equiv \cr &  \equiv \int D[\phi]e^{-\SS_{gr}[\phi ; \rho ,W]} 
\sim 
e^{-\SS_{gr}[\phi_{cl} ; \rho ,W]}.}} 
A crucial assumption in this derivation is that \boucon\ is not a boundary 
condition but a boundary equation of motion, i.e. the measure $D[\phi]$ does not 
depend on $\rho$. 

To summarize the computation, we found that at a finite cutoff the gravity 
action dual to the CFT deformed as in \dfmcft\ is: 
\eqn\newgrac{\SS_{gr}[\phi ; \rho ,W] \equiv \SS_{gr}[\phi ; \rho] 
+W[-\int_{\partial}d^dx'\epsilon^{-\Delta} 
{\bf A}(\vec{x},\vec{x}')\phi (\vec{x}',z)].}
From here we proceed to find the boundary relation in the same way as for 
\bndrsrc: 
\eqn\linrg{\eqalign{\delta\SS_{gr}[\phi ; \rho,W]&=\int_{\BB} d^{d+1}x 
\sqrt{g}\delta\phi[-\nabla^2 +m^2]\phi \cr & 
-\int_{\partial} \epsilon^{-\Delta}\delta\phi 
\left( z^{\Delta-d}(z\partial_z-\Delta)\phi-{\bf A}\rho+{\bf A} 
{\delta W[\psi] \over \delta \psi}\right),}} where we 
denoted the argument of $W$ by $\psi$ which in the condensed notation 
introduced before is given by: 
\eqn\psaie{\psi \equiv-\int_{\partial}d^dx' 
\epsilon^{-\Delta}{\bf A}(\vec{x},\vec{x}')\phi (\vec{x}',z) 
\equiv -\epsilon^{-\Delta}{\bf A}\phi .} 
Since the path integral in \drvtn\ has no boundary condition, a classical 
solution must obey, apart from the bulk equation of motion, also the following 
boundary equation of motion: 
\eqn\newbou{ z^{\Delta-d}(z\partial_z-\Delta)\phi |_{\partial}= 
{\bf A}\left( \rho-{\delta W[\psi] \over \delta \psi} \right).} 
This is the main result of this work. It shows how the deformation of 
the dual CFT by a Multi-trace operator changes the action and the boundary 
equation of motion in gravity.

The generalization to a function of many variables 
$W[\hat{\OO}_1,\dots,\hat{\OO}_n]$ is straightforward. The action becomes:
\eqn\newgrd{\SS_{gr}[\phi_1,\dots,\phi_n;\rho ,W] \equiv \sum_{i=1}^n 
\SS_{gr}[\phi_i ; \rho_i] +W[\psi_1,\dots,\psi_n],} where $\psi_i 
\equiv-\epsilon^{-\Delta_i}{\bf A}_i\phi_i$ and with the following boundary 
equation of motion for each field:  
\eqn\newbg{z^{\Delta_i-d}(z\partial_z-\Delta_i)\phi_i |_{\partial}= 
{\bf A}_i\left( \rho_i-{\delta W[\psi_1,\dots,\psi_n] \over \delta \psi_i} 
\right).} 

In the specific example discussed in \bss:
\eqn\wofbs{W[\hat{\OO}]={\tilde{h} \over 2}\hat{\OO}^2,} we get the 
following deformed boundary equation of motion: 
\eqn\bsbou{z^{\Delta-d}(z\partial_z-\Delta)\phi|_{\partial}= 
{\bf A} \left( \rho +\tilde{h}\epsilon^{-\Delta} 
{\bf A}\phi \right) ,} reproducing 
the result of \bss\ which was obtained using an auxiliary field method. 

In \witmul\ Witten reasoned based on a matrix model analogy that the boundary 
relation in $AdS$ gravity corresponding to a Multi-trace deformation of the 
form \dfmcft\ is 
\eqn\witbou{\alpha={\delta W[\beta] \over \delta \beta},} with $\alpha, \beta$ 
given by \asympt. This matches with the results presented above since 
when the cutoff is removed the LHS of \newbou\ 
essentially picks out the $\alpha$ up to a coefficient (see \bcofuy) 
while from \leema\ the RHS of \newbou\ becomes\foot{This is so because when 
$\epsilon \rightarrow 0$ $\psi \rightarrow \epsilon^{d-2\Delta} \alpha(\vec{x}) 
+ \beta(\vec{x})$ (up to coefficients) and thus ${\delta\psi(\vec{x}) 
\over \delta\beta(\vec{y})} =\delta^d(\vec{x}-\vec{y})$. The extra $\rho$ 
in \newbou\ is just a matter of difference in conventions with respect to 
\witmul\ since we singled out the linear term in $W$.} 
${\delta W[\psi] \over \delta \psi}={\delta W[\beta] \over \delta \beta}$.

\newsec{Discussion and Summary} 

\subsec{Marginality and General Covariance.} 

If $W[\hat{\OO}_1,\dots,\hat{\OO}_n]$ is a marginal deformation in the CFT then 
it contains only monomial terms of dimension $d$. 
Correspondingly $W[-\epsilon^{-\Delta_1}{\bf A}_1\phi_1, \dots , 
-\epsilon^{-\Delta_n}{\bf A}_n\phi_n]$ on the gravity side will have only terms 
of the form $\epsilon^{-d}\prod\phi_j$. In Poincare coordinates the 
boundary volume element is $\sqrt{\widehat{g}}=\epsilon^{-d}$. Thus the 
deformation of the gravity action can be written in a general covariant way. 
From \clasol\ we see that this boundary volume element is needed in order to get 
a finite boundary integral in the limit $\epsilon \rightarrow 0$. 

\subsec{Renormalizability.} 

The way in which the deformation term $W$ enters both the CFT action \dfmcft\ 
and the dual gravity action \newgrac\ suggests that any renormalization 
procedure applied on 
the CFT side by adding local counter-terms translates straightforwardly to the 
gravity side. Thus if the CFT action must be supplemented with a counter-term 
$I+W[\hat{\OO}] \rightarrow I+(W[\hat{\OO}]+R[\hat{\OO}])$ then following the 
same steps as in \drvtn\ we see that the gravity action acquires a corresponding 
boundary term $\SS{gr}+W[\psi] \rightarrow \SS{gr}+(W[\psi]+R[\psi])$. This fact 
was demonstrated explicitly in \bss\ for the case \wofbs. 
  
\subsec{Boundary condition vs boundary equation of motion.} 

An important theme discussed in this note is the issue of boundary condition 
versus boundary equation of motion in $AdS/CFT$. 
In the usual correspondence described in section 2 it seems that there is not 
much of a difference whether one chooses Dirichlet boundary condition or 
\boucon. Furthermore, in the classical limit it is more a matter of semantics 
whether one thinks of \mixed\ as a boundary condition constraining the phase 
space of allowed quantum fluctuations or as a boundary equation of motion 
standing on an equal footing with the bulk equation of motion. 
It is therefore interesting that when one considers Multi-trace deformations the 
natural choice turns out to be \boucon. The derivation \drvtn\ clearly 
supports the interpretation of \boucon\ as a boundary equation of motion and not 
as a boundary condition. The work done in \bss\ also supports this 
interpretation, since although $AdS/CFT$ was formulated there with a boundary 
condition, the result was that after a Double-trace deformation one got 
\newgrac\ without a 
boundary condition\foot{The boundary conditions where explicitly integrated 
over in \bss.}. 
It thus looks more natural if the ``usual" case is formulated in the same way as 
the one with Multi-trace deformations. 
We conclude that our analysis seems to give non-trivial information 
about the quantum gravity side of $AdS/CFT$.
Of course, the claim that $AdS/CFT$ ``has no boundary condition" raises some 
questions regarding conservation of charges etc. 
Also, it is not entirely clear if the distinction is really so sharp.   
Indeed the only fluctuations around the classical 
solution which contribute to the path integral are the ones which behave near 
the boundary like $\delta\phi\sim z^{\kappa}$ with $\kappa > {d \over 2}$ or 
$\kappa = {d \over 2}-\sqrt{{d^2 \over 4}+m^2}$. At least when $\Delta > {d 
\over 2}$ 
fluctuations of the first kind 
automatically satisfy \boucon\ when $\epsilon=0$, while those of the second kind 
are infinitely supressed in the path integral. It thus seems that the difference 
between the two options before the deformation vanishes in this case when one 
removes the cutoff\foot{When $\epsilon =0$ the depth of the well around the 
boundary behavior of the classical solution becomes infinite.}. 
A last remark is that the string theory description of spacetimes with 
boundaries is an interesting and subtle problem; see e.g. the discussion in 
\bss.

\vskip 1cm 

\centerline{\bf Acknowledgments} 

We would like to thank R. Argurio, J.L.F. Barbon, D.S. Berman, M. Dine, S. 
Elitzur, G. Ferretti, 
A. Giveon, A. Hanany, Y. Oz, A. Petkou, E. Rabinovici for useful discussions. 
We owe special thanks to M. Berkooz and O. Aharony for advising us and for a 
critical reading of the manuscript. 
A.Sh thanks the theory devision in CERN and the string group in Chalmers 
university, Gothenburg, for hospitality during the final stages of this work. 
A.Sh is supported by a Clore fellowship. 
This work is supported in part by BSF -- American-Israel Bi-National 
Science Foundation, the Israel Academy of Sciences and Humanities --
Centers of Excellence Program, the German-Israel Bi-National Science 
Foundation, the European RTN network HPRN-CT-2000-00122.

\listrefs   
   
\end